\documentclass{article}

\usepackage{arxiv}

\usepackage[utf8]{inputenc} 
\usepackage[T1]{fontenc}    
\usepackage{hyperref}       
\hypersetup{
  colorlinks   = true, 
  urlcolor     = red, 
  linkcolor    = red, 
  citecolor   = green 
}
\usepackage{url}            
\usepackage{booktabs}       
\usepackage{amsfonts}       
\usepackage{nicefrac}       
\usepackage{microtype}      
\usepackage{subfigure} 

\usepackage{framed,multirow}

\usepackage{amsmath}
\usepackage{algorithm}
\usepackage{algpseudocode}
\usepackage{graphicx}

\usepackage{amssymb}
\usepackage{latexsym}
\usepackage{bm}
\usepackage{amssymb}
\usepackage{array}
\usepackage{multirow}
\usepackage[english]{babel}
\usepackage[numbers]{natbib}
\usepackage{footnote}
\usepackage{tabularx}
\makesavenoteenv{tabular}
\makesavenoteenv{table}
\usepackage{cleveref}
\crefname{equation}{Eq.}{Eqs.}
\crefname{table}{Table}{Tables}
\crefname{figure}{Fig.}{Figs.}
\crefname{section}{Section}{Secs.}
\crefname{subsection}{Section}{Secs.}
\Crefname{figure}{Figs.}{Figs.}
\Crefname{Algorithm}{Algorihtm}{Algorihtm}
\usepackage{xcolor}
\definecolor{newcolor}{rgb}{.8,.349,.1}
\title{A Direct Hamiltonian MCMC Approach for Reliability Estimation}

\author{
  Hamed~Nikbakht\\
  Department of Civil and Environmental Engineering\\
  The Pennsylvania State University\\
  University Park, PA 16802 \\
  \texttt{hun35@psu.edu} \\
   \And
 Konstantinos G.~Papakonstantinou \\
  Department of Civil and Environmental Engineering\\
  The Pennsylvania State University\\
  University Park, PA 16802 \\
}
\begin{document}
\maketitle

\begin{abstract}
Accurate and efficient estimation of rare events probabilities is of significant importance, since often the occurrences of such events have widespread impacts. The focus in this work is on precisely quantifying these probabilities, often encountered in reliability analysis of complex engineering systems, by introducing a gradient-based Hamiltonian Markov Chain Monte Carlo (HMCMC) framework, termed Approximate Sampling Target with Post-processing Adjustment (ASTPA). The basic idea is to construct a relevant target distribution by weighting the high-dimensional random variable space through a one-dimensional likelihood model, using the limit-state function. To sample from this target distribution we utilize HMCMC algorithms that produce Markov chain samples based on Hamiltonian dynamics rather than random walks. We compare the performance  of  typical  HMCMC  scheme  with our newly developed Quasi-Newton  based  mass preconditioned HMCMC algorithm that can sample very adeptly, particularly in difficult cases with high-dimensionality and very small failure probabilities. To eventually compute the probability of interest, an original post-sampling step is devised at this stage, using an inverse importance sampling procedure based on the samples. The involved user-defined parameters of ASTPA are then discussed and general default values are suggested. Finally, the performance of the proposed methodology is examined in detail and compared against Subset Simulation in a series of static and dynamic low- and high-dimensional benchmark problems.
\end{abstract}


\section{INTRODUCTION}
In this work, we investigate Hamiltonian Markov Chain Monte Carlo (HMCMC) schemes for estimation of rare events probabilities, a commonly encountered important problem in several engineering and scientific applications \citep{Nikbakht2011loading,Nikbakht2013comparison}, most often observed in the form of failure probability, or alternatively, reliability estimation. Calculating such small probabilities with accuracy presents many numerical and mathematical challenges, particularly in cases with high dimensional random spaces and/or expensive computational models, that practically limit the afforded number of model calls. The well known gradient based First Order Reliability Method (FORM), and variants, have a very long history in reliability estimation problems, with numerous successes \citep{rackwitz2001reliability,der2005first,liu1991optimization,breitung201540}. Such asymptotic approximation methods naturally have of course limitations, however, in general settings. Hence, numerous sampling based methods have been also suggested in the literature to tackle the problem in its utmost generality, e.g. \citep{schueller2007benchmark}. The current state-of-the-art sampling method for problems of this type is termed Subset Simulation (SuS) \citep{au2001estimation} and belongs to the family of MCMC techniques. Within the context of Subset Simulation, various random-walk and non-random-walk-based MCMC proposal steps \citep{papaioannou2015mcmc,zuev2015subset} have been explored and suggested, to improve the sampling efficiency of SuS, including Hamiltonian steps \citep{wang2019hamiltonian}.\par
In this work we completely deviate from SuS and we introduce a gradient-based Hamiltonian Markov Chain Monte Carlo (HMCMC) sampling framework, termed \textit{Approximate Sampling Target with Post-processing Adjustment} (ASTPA) \citep{Nikbakht2019HMCMC}, that is directly used for rare  events probabilities estimation. The basic idea of ASTPA is to construct a relevant target distribution to sample from, by weighting the high-dimensional random variable space through a one-dimensional likelihood model, using the limit-state function, and to then utilize an  original post-sampling step, using an inverse importance sampling procedure based on the acquired samples. Hamiltonian MCMC schemes are employed  to perform the sampling. The Hamiltonian Monte Carlo (HMC) method, originally developed by \citep{duane1987hybrid}, and more recently popularized mainly through the works of \citep{neal2012bayesian,neal2011mcmc,hoffman2014no,girolami2011riemann}, is characterized by scalability \citep{neal2011mcmc,beskos2013optimal,girolami2011riemann}, fast mixing rates, weak sample auto-correlation, even in complex high-dimensional parameter spaces \citep{betancourt2017conceptual,kamani2016shape,kamani2018skeleton}, and has achieved broad-spectrum successes in most general settings e.g. \cite{gelman2013bayesian,kruschke2014doing,monnahan2017faster,akhmatskaya2008gshmc}. Herein, we compare the performance of the typical HMCMC scheme with our newly developed Quasi-Newton based mass preconditioned HMCMC algorithm that also exploits the information about the localized geometry of the failure region, through an inexpensive BFGS approximation. The involved user-defined parameters of ASTPA are also discussed in the paper and general default values are suggested. The performance of the proposed methodology is finally examined and compared successfully against Subset Simulation, in a series of static and dynamic, low- and high-dimensional benchmark problems.
\section{CONCEPTS BEHIND HAMILTONIAN MARKOV CHAIN MONTE CARLO}\label{section2}
In HMCMC methods, Hamiltonian dynamics are used to produce distant state steps for the Metropolis proposals, thereby avoiding the slow exploration of the state space that results from the diffusive behavior of simple random-walk proposals. Given a parameter of interest $\boldsymbol{\theta}$ with (unnormalized) density $\pi_{\Theta}$(.), the Hamiltonian Markov Chain Monte Carlo method introduces an auxiliary momentum variable $\textbf{z}$ and samples from the joint
distribution characterized by:
\begin{align} \label{eq:8}
\pi(\boldsymbol{\theta},\textbf{z}) \propto  \pi_{\Theta}(\boldsymbol{\theta})\ \pi_{Z \arrowvert\Theta}(\textbf{z} \arrowvert \boldsymbol{\theta})
\end{align}   
where $\pi_{Z\arrowvert \Theta}(.\arrowvert \boldsymbol{\theta})$ is proposed to be a symmetric distribution. With $\pi_{\Theta}(\boldsymbol{\theta})$ and $\pi_{Z\arrowvert \Theta}(\textbf{z}\arrowvert \boldsymbol{\theta})$ being uniquely described up to normalizing constants, the functions $U(\boldsymbol{\theta}) = -\log\pi_{\Theta}(\boldsymbol{\theta})$ and $K(\boldsymbol{\theta},\textbf{z})=-\log\pi_{Z\arrowvert \Theta}(\textbf{z}\arrowvert \boldsymbol{\theta})$ are introduced as the potential energy and kinetic energy, owing to the physical laws which motivate the Hamiltonian Markov Chain Monte Carlo algorithm. The total energy $H(\boldsymbol{\theta},\textbf{z})$ can be thus expressed as:
\begin{align} \label{eq:Hamiltonian}
H(\boldsymbol{\theta},\textbf{z})=U(\boldsymbol{\theta})+K(\boldsymbol{\theta},\textbf{z}) 
\end{align}   
and is often termed the Hamiltonian $H$. The kinetic energy function is unconstrained and can be formed in various ways based on the implementation. In most typical cases, the momentum is given by a zero-mean normal distribution \citep{neal2011mcmc,betancourt2017conceptual}, and accordingly the kinetic energy can be written as: $K(\boldsymbol{\theta},\textbf{z}) = -\log\pi_{Z\arrowvert \Theta}(\textbf{z}\arrowvert \boldsymbol{\theta})=-\log\pi_{Z}(\textbf{z}) = \frac{1}{2}\textbf{z}^{T} \textbf{M}^{-1} \textbf{z}$, where the $\textbf{M}$ is a symmetric, positive-definite covariance (mass) matrix. \par
HMCMC generates a Metropolis proposal on the joint state-space $(\boldsymbol{\theta},\textbf{z})$ by sampling the momentum and simulating trajectories of Hamiltonian dynamics in which the time evolution of the state $(\boldsymbol{\theta},\textbf{z})$ is governed by Hamilton's equations, expressed typically by: 
\begin{equation}
\frac{d\boldsymbol{\theta}}{dt} = \frac{\partial H}{\partial \textbf{z}} =\frac{\partial K}{\partial \textbf{z}}= \textbf{M}^{-1} \textbf{z} ,\ \ \  \ 
\frac{d \textbf{z}}{dt} = -\frac{\partial H}{\partial \boldsymbol{\theta}} = -\frac{\partial U}{\partial \boldsymbol{\theta}}= \nabla_{\theta} \mathcal{L}(\boldsymbol{\theta}) \label{eq:10}
\end{equation}  
where $\mathcal{L}(\boldsymbol{\theta})$ denotes the log-density of the target distribution. Hamiltonian dynamics prove to be an effective proposal generation mechanism because the distribution 
$\pi(\boldsymbol{\theta},\textbf{z})$ is invariant under the dynamics of \cref{eq:10}. These dynamics enable a proposal state, obtained by an
approximate solution of \cref{eq:10}, to be distant from the current state, yet having high probability of acceptance. The solution to \cref{eq:10} is in general analytically intractable and thus the Hamiltonian equations need to be numerically solved by discretizing time, using some small step size, $\varepsilon$. A symplectic integrator that can be used for the numerical solution is the leapfrog one, as follows:
\begin{equation} 
\textbf{z}\textsubscript{t+$\varepsilon$/2} = \textbf{z}\textsubscript{t} - (\dfrac{\varepsilon}{2})\dfrac{\partial U}{\partial \boldsymbol{\theta}} (\boldsymbol{\theta}\textsubscript{t}) \label{eq:11} ,\ \ \  \
\boldsymbol{\theta}\textsubscript{t+$\varepsilon$} = \boldsymbol{\theta}\textsubscript{t} + \varepsilon \dfrac{\partial K}{\partial \textbf{z}} (\textbf{z}\textsubscript{t+$\varepsilon$/2}) ,\ \ \  \
\textbf{z}\textsubscript{t+$\varepsilon$} = \textbf{z}\textsubscript{t+$\varepsilon$/2} - (\dfrac{\varepsilon}{2})\dfrac{\partial U}{\partial \boldsymbol{\theta}} (\boldsymbol{\theta}\textsubscript{t+$\varepsilon$}) 
\end{equation}
The main advantages of using the leapfrog integrator are its simplicity, its volume-preserving feature, and its reversibility, due to its symmetry, by simply negating $\textbf{z}$, facilitating a valid Metropolis proposal. See \citep{neal2011mcmc}, \citep{betancourt2017conceptual} and\citep{tripuraneni2016magnetic} for details on energy-conservation, reversibility and volume-preserving integrators and their connections to HMCMC. It is noted that in the above leapfrog integration algorithm, the computationally expensive part is to acquire the $\dfrac{\partial U}{\partial \boldsymbol{\theta}}$ term at the updated location $\boldsymbol{\theta}$. Taking $\textit{L}=\tau/\varepsilon$ steps of the leapfrog integrator approximates the evolution $(\boldsymbol{\theta}(0),\textbf{z}(0)) \longrightarrow (\boldsymbol{\theta}(\tau),\textbf{z}(\tau))$, where $\tau$ is the trajectory length or
path length, and provides the exact solution in the limit $\varepsilon \longrightarrow 0$.\par
\begin{algorithm}[t!]
\caption{Hamiltonian Markov Chain Monte Carlo}\label{alg:stdHMCMC}
\begin{algorithmic}[1]
\Procedure{HMCMC}{$\boldsymbol{\theta}^{0}$, $\varepsilon$, \textit{L}, $\mathcal{L}(\boldsymbol{\theta})$, \textit{NIter}}
\For{\texttt{$m=1$ $to$ $NIter$}}
\State $\textbf{z}^{0}$$\sim$$N(0,\textbf{I})$\Comment{momentum sampling from standard normal distribution}
\State $\boldsymbol{\theta}^{m}$ $\gets$ $\boldsymbol{\theta}^{m-1}$, $\tilde{\boldsymbol{\theta}}$ $\gets$ $\boldsymbol{\theta}^{m-1}$, $\tilde{\textbf{z}}$ $\gets$ $\textbf{z}^{0}$
\For{\texttt{$i=1$ $to$ $L$}}
\State $\tilde{\boldsymbol{\theta}}$, $\tilde{\textbf{z}}$ $\gets$ Leapfrog($\tilde{\boldsymbol{\theta}}$, $\tilde{\textbf{z}}$, $\varepsilon$) \Comment{leapfrog integration}
\EndFor\label{HMCMCfor2}
\State $with$ $probability$:\\ 
       \hspace{1cm} 
       $\alpha$ = min$\bigg\{$1,$\dfrac{\exp(\mathcal{L}(\tilde{\boldsymbol{\theta}})-\dfrac{1}{2} \tilde{\textbf{z}}.\tilde{\textbf{z}})}{\exp(\mathcal{L}(\boldsymbol{\theta}^{m-1})-\dfrac{1}{2}\textbf{z}^{0}.\textbf{z}^{0})}$
       $\bigg\}$ \Comment{Metropolis step}\\
       \hspace{1cm} $\boldsymbol{\theta}^{m}$ $\gets$ $\tilde{\boldsymbol{\theta}}$, $\textbf{z}^{m}$ $\gets$ -$\tilde{\textbf{z}}$
\EndFor\label{HMCMCfor}
\EndProcedure
\end{algorithmic}
\end{algorithm}
As discussed, the typical HMCMC version is based on a Gaussian momentum $\pi_{Z \arrowvert\Theta}(\textbf{z} \arrowvert \boldsymbol{\theta}) = \pi_{Z}(\textbf{z}) \sim N(\textbf{0},\textbf{M})$ (or $\textbf{z} \sim N(\textbf{0},\textbf{M})$). The mass matrix $\textbf{M}$ is often set to the identity matrix, \textbf{I}, but can also be adapted to precondition the sampler when relevant information about the target distribution is available (see \cref{section3}). A standard procedure for drawing \textit{NIter} samples via HMCMC is described in \Cref{alg:stdHMCMC}, where $\mathcal{L}(\boldsymbol{\theta})$ is the log-density of the target distribution of interest. $\boldsymbol{\theta}^{0}$  are the initial values for the \textit{$\boldsymbol{\theta}$}, and \textit{L} is the number of leapfrog steps, as explained before. For each HMCMC step, we first resample the momentum and then implement the \textit{L} leapfrog updates (Leapfrog($\tilde{\boldsymbol{\theta}}$, $\tilde{\textbf{z}}$, $\varepsilon$)) before we accept or reject the Metropolis proposal at the pertinent step.\par
The efficiency of HMCMC relies significantly on selecting suitable values for $\varepsilon$ and \textit{L}. In this work we select the stepsizes $\varepsilon$ in such a way that the corresponding average acceptance rates are approximately 65$\%$, as values between 60$\%$ and 80$\%$ are typically assumed optimal \citep{neal2011mcmc,hoffman2014no,beskos2013optimal}. The dual averaging algorithm of \citep{hoffman2014no} was adopted here to find these stepsizes. To determine the value of $L$, we estimate the trajectory length $\tau$ so as to have a sufficient so called normalized Expected Square Jumping Distance (\textit{ESJD}) $\tau^{-1/2}\mathop{\mathbb{E}}$\(\lVert\theta^{(t+1)} (\tau) - \theta^{(t)} (\tau)\rVert\)$^{2}$, as introduced in \citep{wang2013adaptive}, and then we randomly perturb each trajectory length $\tau^{(t)}$ in the range $[0.9\tau,1.1\tau]$ to avoid periodicity ($t$ denotes the $t$-th iteration of HMCMC). In all our experiments we determine $L$ and control the trajectory length in this manner, as we have found it to work well in practice. The role of these parameters ($\varepsilon$ and $\tau$  (or $L$)) and techniques for determining them have been quite extensively studied and for more details we refer the readers to \citep{neal2011mcmc,hoffman2014no,beskos2013optimal}.
\section{METHODOLOGY TO CALCULATE THE FAILURE PROBABILITY}\label{section4}
The failure probability \textit{P\textsubscript{F}} for a system, that is the probability of a defined unacceptable system performance, can be expressed as a $d$-fold integral, as:
\begin{equation} \label{eq:1}
\textit{P\textsubscript{F}}= \mathop{\mathbb{E}}[I_{F} (\boldsymbol{\theta})] = \int_{g(\boldsymbol{\theta})\leq 0} I_{F} (\boldsymbol{\theta}) \pi_{\boldsymbol{\theta}}(\boldsymbol{\theta}) d\boldsymbol{\theta} 
\end{equation} 
where \boldsymbol{$\theta$} is the random vector $[\theta_{1},..., \theta_{d}]$$^{T}$ ; $F \subset \mathbb{R}^{d}$ is the failure event in the parameter space; g(\textbf{\boldsymbol{$\theta$}}) is the limit-state function that can include one or several distinct failure modes and defines the failure of the system by g(\textbf{\boldsymbol{$\theta$}})$\leq$ 0; \textit{I(.)} is the indicator function with: \textit{$I_{F}$} (\boldsymbol{$\theta$}) = 1 if \boldsymbol{$\theta$} $\in$ g(\textbf{\boldsymbol{$\theta$}})$\leq$ 0 and $I_{F}$(\boldsymbol{$\theta$}) = 0 otherwise; $\mathop{\mathbb{E}}$ is the expectation operator, and $\pi_{\boldsymbol{\theta}}$ is the joint probability density function (PDF) for \textbf{$\Theta$}. It is common practice in reliability analysis to have the joint PDF of \textbf{$\Theta$} be  the standard normal one, due to its rotational symmetry and exponential probability decay. In most cases, this is not restrictive, since it is uncomplicated to transform the original random variables \textbf{X} to \textbf{$\Theta$}, e.g. \citep{hohenbichler1981non}. When this is not the case however, but the probabilistic characterization of  \textbf{X} can be defined in terms of marginal distributions and correlations, the Nataf distribution (equivalent to Gaussian copula) can be used to model the joint PDF, and the mapping to the standard normal space can be then accomplished \citep{der1986structural}.\par
The main idea of our approach to calculate the failure probability is to construct an appropriate approximate target distribution to sample from, based on Hamiltonian MCMC methods that can quickly reach regions of interest and can keep the number of model calls to a minimum, and to then utilize a post-sampling step to acquire the exact probability estimation, without any additional model calls. We construct this approximate target distribution by combining the multidimensional parameter space $\Theta$ with a one-dimensional likelihood function, using the limit-state expression.  
\begin{figure}[t!]
 \centering
  \begin{tabular}{c ccccc}
   \includegraphics[width=.3\textwidth,keepaspectratio]{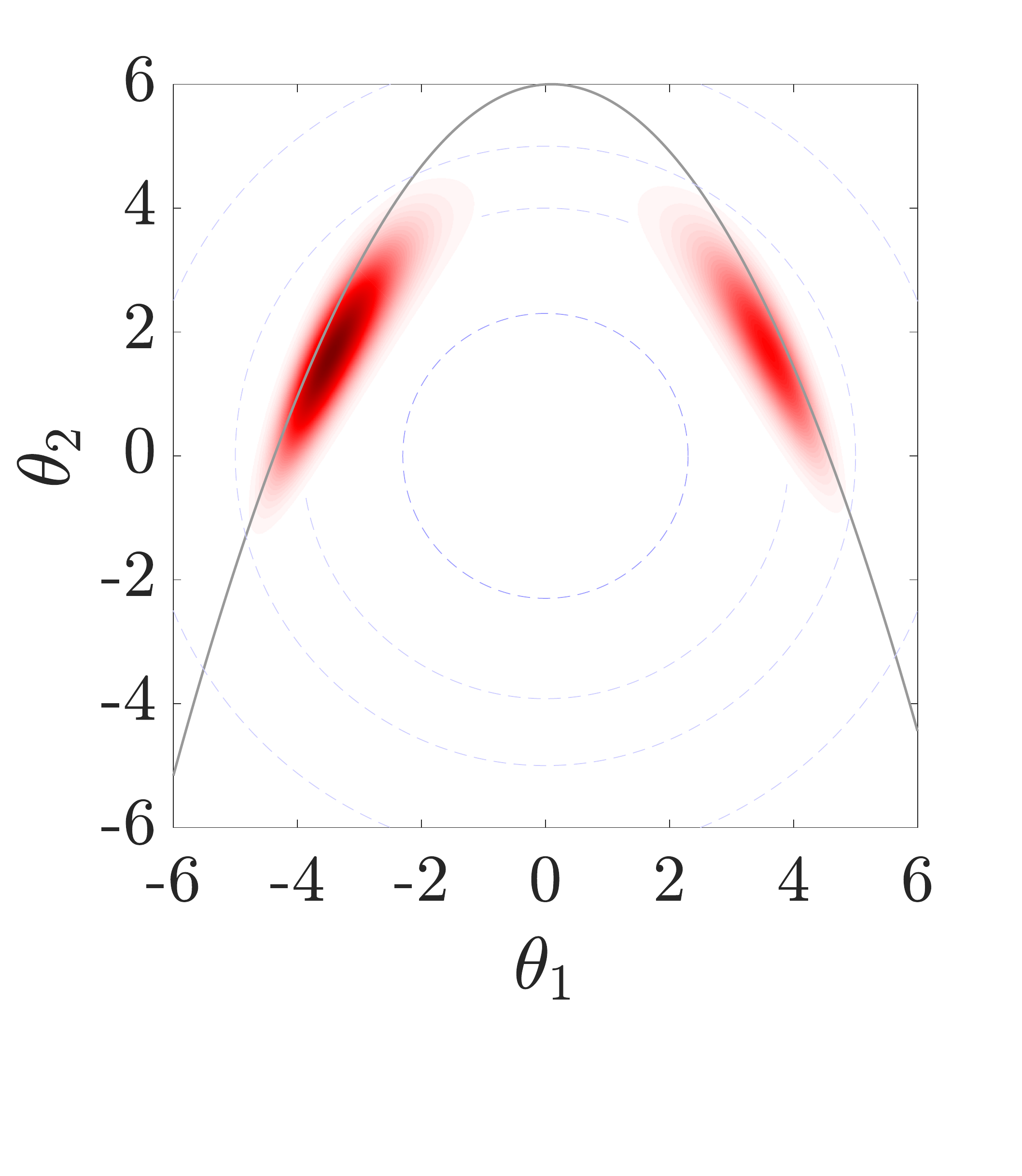}&
   \includegraphics[width=.3\textwidth,keepaspectratio]{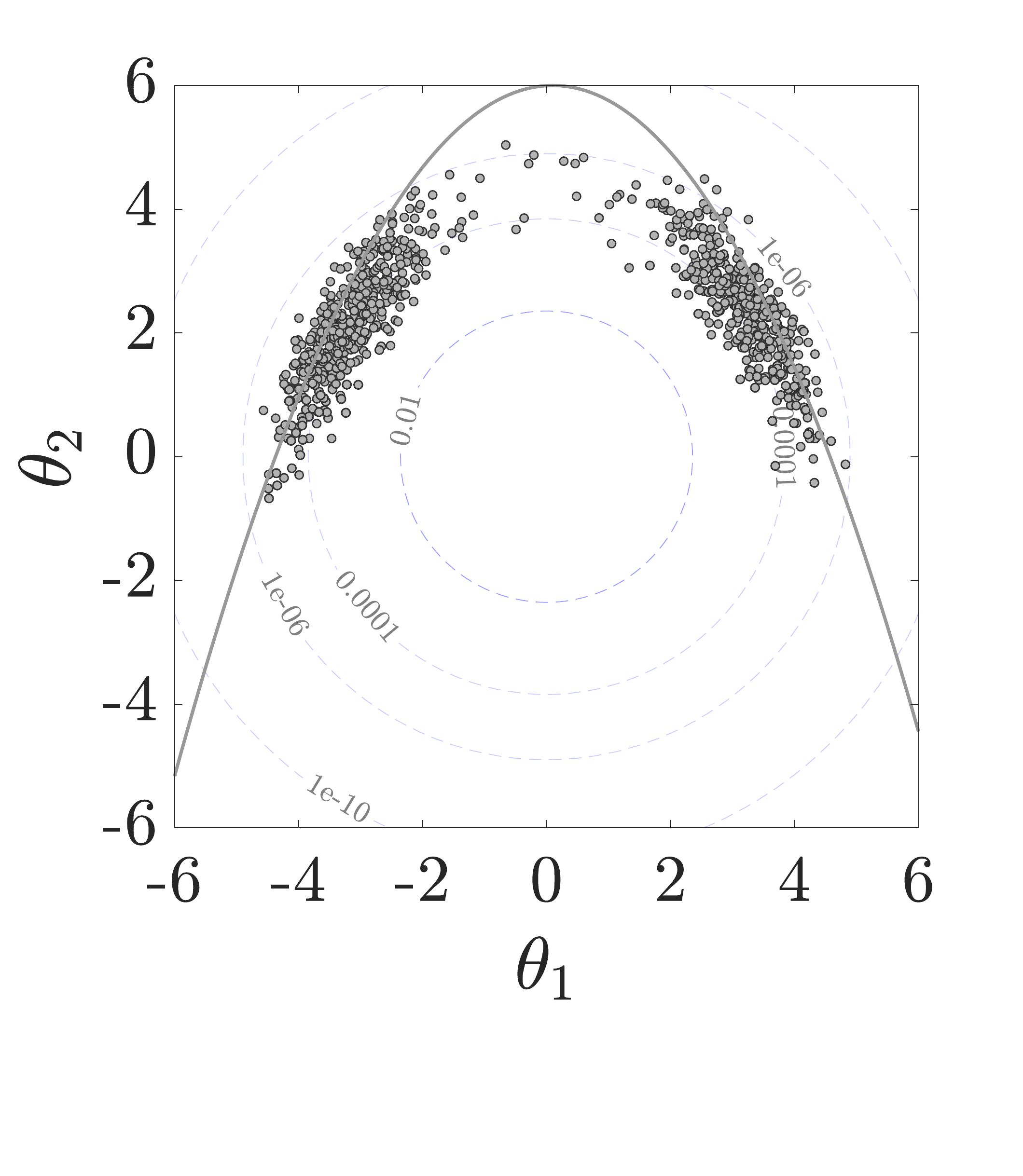}&
    \includegraphics[width=.3\textwidth,keepaspectratio]{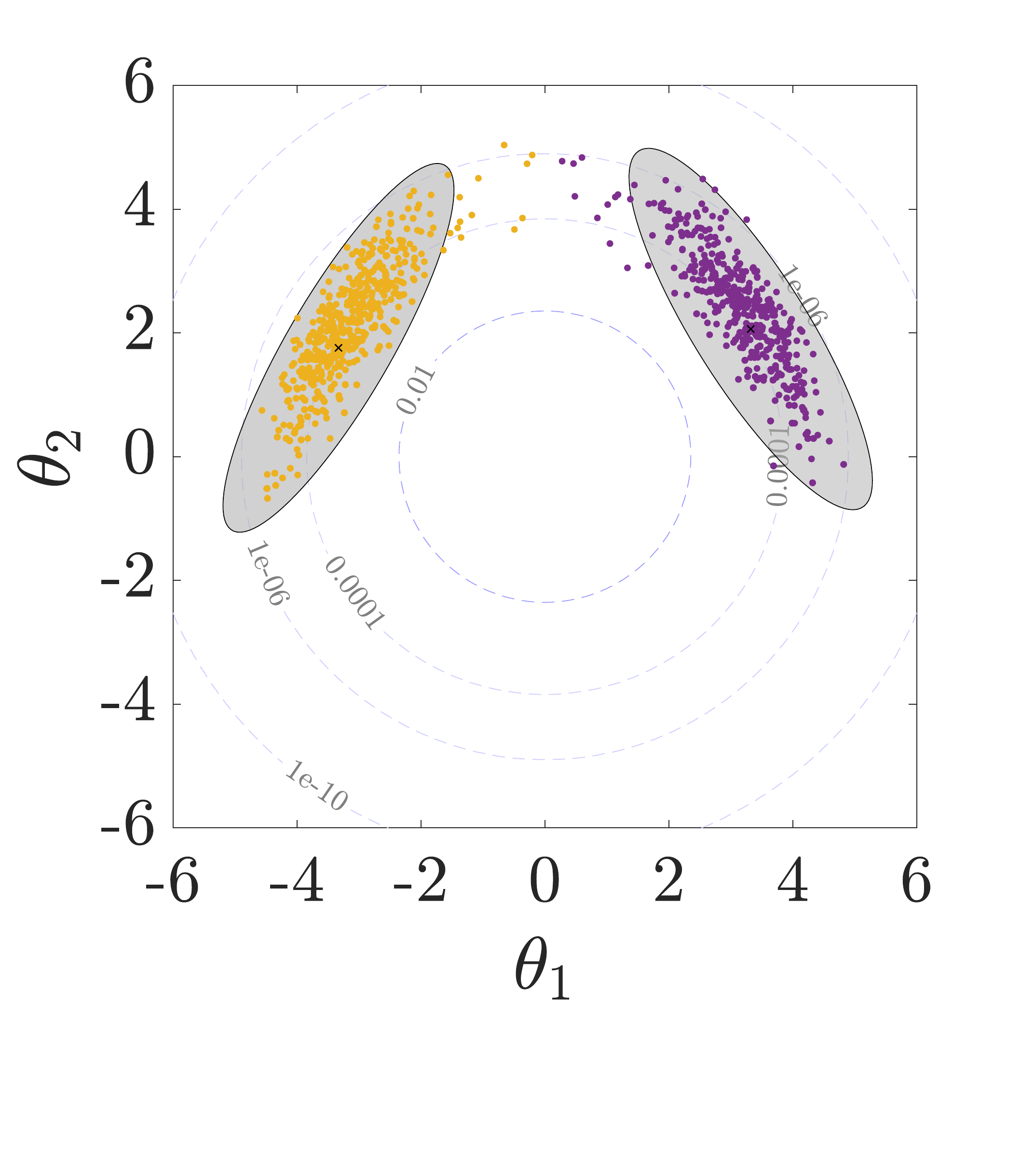}
    \vspace*{-0.3in}
\\\hline\noalign{\smallskip}

 \end{tabular}
 \caption{The above figures represent the analytical target distribution, the simulated target distribution samples based on our HMCMC-based method, and the fitted Gaussian Mixture Model describing the simulated samples, from left to right, respectively.}
 \label{fig2}
\end{figure}
\noindent This one-dimensional likelihood function is expressed as a Gaussian PDF with mean $= \mu_{g(\boldsymbol\theta)} = 0$, where $g(\boldsymbol\theta)$ is the limit-state function, and a dispersion factor $ \sigma$:
\begin{align}
\textbf{N}\bigg (\dfrac{g(\boldsymbol\theta)}{g_{c}}\bigg\arrowvert\ \mu_{g(\boldsymbol\theta)}=0 , \ \sigma\bigg ),\ \ \ \ 
g_{c}=
\begin{cases}
g(\textbf{0}),& \text{if}  \, g(\textbf{0})>8  \, \text{or}  \, g(\textbf{0})<1\\
1,    & \text{otherwise}
\end{cases}\label{eqq8}
\end{align}
where $g_{c}$ is a normalizing constant. The reason for this normalization, $g(\boldsymbol\theta)/g_{c}$, is to control the suggested upper and lower bounds of $\sigma$. The target PDF is then defined as:
\begin{align}
{\text{Target probability distribution}}\propto \textbf{N}\bigg (\dfrac{g(\boldsymbol\theta)}{g_{c}}\bigg\arrowvert\ \mu_{g(\boldsymbol\theta)}=0 , \ \sigma\bigg )\times \bigg (\boldsymbol\theta\sim \textbf{N}(\textbf{0},\textbf{I})\bigg )
\end{align}
Having the total number of model calls in mind, as well as the coefficient of variation of the estimator (C.O.V), the suggested value for $\sigma$ is in the range $[0.1 \, \,0.7]$. Fine tuning $\sigma$ in that range is not generally necessary. It is recommended, in general, to use higher $\sigma$ values $(0.6-0.7)$ in nonlinear high-dimensional problems and multi-modal cases when $1\leq g(\textbf{0}) \leq8$, since a larger $\sigma$ usually allows longer state jumps and fewer required model calls. On the other hand, a lower $\sigma$ generally increases the accuracy of the estimator, at the expense of a slightly increased number of model calls.\par   
\cref{fig2} concisely portrays the overall approach by using a bimodal target distribution. The gray curves represent the parabolic limit-state function $g(\boldsymbol{\theta})$ of this problem, with the failure domain being outside $g(\boldsymbol{\theta})$. The left figure displays the constructed target distribution, by adopting the previously described approach, which in this simple 2D case can be visualized. The middle figure shows drawn samples from the target distribution by our suggested Hamiltonian MCMC variant, described in \cref{section3}. For their initial stage, our HMCMC samplers have an adaptive annealed phase, mainly in order to automatically tune parameters and reduce the computational cost, overall, and then follow the typical Hamiltonian approach, except in our Quasi-Newton case (\cref{section3}) the mass matrix is appropriately preconditioned. As such, during the burn-in period, we initialize the spread of likelihood, $\sigma_{0}$, equal to $1$ that then follows an \textit{exponential decay} throughout the burn-in period, while at the end of this initial period, $\sigma$ takes its constant value, as described above, for the stationary phase of the algorithms.\par 
To finally compute the failure probability we have to adjust \cref{eq:1} accordingly, since the samples have been sampled based on our constructed approximate target distribution. An original post-sampling step is devised at this stage using our inverse importance sampling procedure, i.e. having the samples, choose a pertinent Importance Sampling Density (ISD) automatically, based on the samples. Given that, the probability of failure after some algebra (see \citep{Nikbakht2019HMCMC} for details) can be computed as follows:  
\begin{align}    
\textit{P\textsubscript{F}} = \int I_{F} (\boldsymbol{\theta}) \pi_{\boldsymbol{\theta}}(\boldsymbol{\theta}) d\boldsymbol{\theta}= \int I_{F}(\boldsymbol{\theta}) \dfrac{C\ .\ \tilde{h}(\boldsymbol{\theta})}{\ell(\boldsymbol{\theta})} d\boldsymbol{\theta}
\end{align}
where $C = \frac{1}{N} \sum_{i=1}^{N} \dfrac{\tilde{h}(\theta_{i})}{Q(\theta_{i})}$; $\tilde{h}(.)$ denotes the non-normalized target PDF, $\ell(\boldsymbol{\theta})$ is our likelihood function, and $Q(.)$ is a computed Gaussian Mixture Model (GMM), based on the already available samples and the generic Expectation Maximization (EM) algorithm, as indicatively seen in the right plot of \cref{fig2}.\par
Our described newly proposed method is termed ASTPA (Approximate Sampling Target with Post-processing Adjustment) and, as a summary, comprises of constructing a target distribution model, performing HMCMC sampling, and finally applying a post-sampling step. For more details on supplementary justifications about this method, we refer readers to \citep{Nikbakht2019HMCMC}.
\section{QUASI-NEWTON EXTENSIONS AND CONNECTIONS TO HMCMC}\label{section3}
In high-dimensional problems, the computational cost of the typical HMCMC sampler may increase considerably and a prohibitive number of model calls per leapfrog step may be required. In this work, we address this issue in a developed Newton-type context,  where the Hessian information is approximated without any required additional model calls per leapfrog step. To this end, the well-known BFGS approximation \citep{nocedal2006numerical} is used in our Quasi-Newton type Hamiltonian MCMC approach. 
Let $\boldsymbol{\theta}\in\mathbb{R}^{d}$, consistent with the previous section. Given the $k$-$th$ estimate $\textbf{W}_{k}$, where $\textbf{W}_{k}$ is an approximation to the inverse Hessian at $\boldsymbol{\theta}_{k}$, the BFGS update $\textbf{W}_{k+1}$ can be expressed as:
\begin{align}
\textbf{W}_{k+1} = (\textbf{I}-\dfrac{\textbf{\textit{s}}_{k} \textbf{\textit{y}}_{k}^{T}}{\textbf{\textit{y}}_{k}^{T} \textbf{\textit{s}}_{k}})\textbf{W}_{k}(\textbf{I}-\dfrac{\textbf{\textit{y}}_{k} \textbf{\textit{s}}_{k}^{T}}{\textbf{\textit{s}}_{k}^{T} \textbf{\textit{y}}_{k}})+\dfrac{\textbf{\textit{s}}_{k} \textbf{\textit{s}}_{k}^{T}}{\textbf{\textit{s}}_{k}^{T} \textbf{\textit{y}}_{k}} \label{eq:14}
\end{align}
where \textbf{I} is the identity matrix, $\textbf{\textit{s}}_{k}=\boldsymbol{\theta}_{k+1}-\boldsymbol{\theta}_{k}$, and $\textbf{\textit{y}}_{k}= \nabla f(\boldsymbol{\theta}_{k+1})-\nabla f(\boldsymbol{\theta}_{k})$ where $f:\mathbb{R}^{d} \longrightarrow\mathbb{R}$ denotes any relevant target distribution function in this case. Our developed Quasi-Newton preconditioned Hamiltonian Markov Chain Monte Carlo (QNp-HMCMC) method is presented in detail in \Cref{QNHMCMC}. In the burn-in phase we are still sampling the momentum from an identity mass matrix but the ODEs of \cref{eq:10} now become:    
\begin{equation}
\dot{\boldsymbol{\theta}}=\textbf{W}\textbf{M}^{-1}\textbf{z},\ \ \  \ 
\dot{\textbf{z}}=\textbf{W}\nabla_{\theta} \mathcal{L}(\boldsymbol{\theta}).\label{eqQN}
\end{equation}
where $\textbf{W}\in\mathbb{R}^{d\times d}$ is the symmetric positive definite matrix of \cref{eq:14} and being the inverse Hessian matrix provides an informed approximation of the local geometry of the parameter space, accelerating exploration of the domain. The final estimation of the approximated inverse of the Hessian matrix, \textbf{W}, from the burn-in phase is then used to define the preconditioned covariance matrix to sample the momentum variable for the stationary, non-adaptive stage of the chain. It can be shown that all utilized dynamics in both phases of the algorithm enable us to maintain the desired target distribution as the invariant one. In \cref{section5} we empirically evaluate and compare the QNp-HMCMC performance in various settings. For further details on the QNp-HMCMC method, its performance in different settings, and its validity, see \citep{Nikbakht2019HMCMC}.
\begin{algorithm}[t!]
\caption{Quasi-Newton preconditioned Hamiltonian Markov Chain Monte Carlo}\label{QNHMCMC}
\begin{algorithmic}[1]
\Procedure{QNp-HMCMC}{$\boldsymbol{\theta}^{0}$, $\varepsilon$, \textit{L}, $\mathcal{L}(\boldsymbol{\theta})$, \textit{BurnIn}, \textit{NIter}}\\
 \hspace{0.5cm}\textbf{W} = \textbf{I}
\For{\texttt{$m=1$ $to$ $NIter$}}
\If {$m$ $\leq$ $BurnIn$}
\State $\textbf{z}^{0}$$\sim$$\textbf{N}(\textbf{0},\textbf{M})$\Comment{where $\textbf{M}=\textbf{I}$}
\State $\boldsymbol{\theta}^{m}$ $\gets$ $\boldsymbol{\theta}^{m-1}$, $\tilde{\boldsymbol{\theta}}$ $\gets$ $\boldsymbol{\theta}^{m-1}$, $\tilde{\textbf{z}}$ $\gets$ $\textbf{z}^{0}$, $\textbf{B}$ $\gets$ $\textbf{W}$
\For{\texttt{$i=1$ $to$ $L$}}
\State $\tilde{\boldsymbol{\theta}}$, $\tilde{\textbf{z}}$ $\gets$ Leapfrog-BurnIn($\tilde{\boldsymbol{\theta}}$, $\tilde{\textbf{z}}$, $\varepsilon$, $\textbf{B}$)\\
\hspace{2cm} Update \textbf{W} using \cref{eq:14}
\EndFor
\State $with$ $probability$:\\ 
       \hspace{1.5cm}$\alpha$ = min$\bigg\{$1,$\dfrac{\exp(\mathcal{L}(\tilde{\boldsymbol{\theta}})-\dfrac{1}{2} \tilde{\textbf{z}}.\tilde{\textbf{z}})}{\exp(\mathcal{L}(\boldsymbol{\theta}^{m-1})-\dfrac{1}{2}\textbf{z}^{0}.\textbf{z}^{0})}$
       $\bigg\}$\\
       \hspace{1.5cm} $\boldsymbol{\theta}^{m}$ $\gets$ $\tilde{\boldsymbol{\theta}}$, $\textbf{z}^{m}$ $\gets$ -$\tilde{\textbf{z}}$\Comment{If proposal rejected: $\textbf{W}$ $\gets$ $\textbf{B}$}
\Else \Comment{If $m$ $>$ $BurnIn$}      
\State $\textbf{z}^{0}$$\sim$$\textbf{N}(\textbf{0},\textbf{M})$\Comment{where $\textbf{M}=\textbf{W}^{-1}$}
\State $\boldsymbol{\theta}^{m}$ $\gets$ $\boldsymbol{\theta}^{m-1}$, $\tilde{\boldsymbol{\theta}}$ $\gets$ $\boldsymbol{\theta}^{m-1}$, $\tilde{\textbf{z}}$ $\gets$ $\textbf{z}^{0}$
\For{\texttt{$i=1$ $to$ $L$}}
\State $\tilde{\boldsymbol{\theta}}$, $\tilde{\textbf{z}}$ $\gets$ Leapfrog($\tilde{\boldsymbol{\theta}}$, $\tilde{\textbf{z}}$, $\varepsilon$, $\textbf{M}$)
\EndFor
\State $with$ $probability$:\\ 
       \hspace{1.5cm}$\alpha$ = min$\bigg\{$1,$\dfrac{\exp(\mathcal{L}(\tilde{\boldsymbol{\theta}})-\dfrac{1}{2} \tilde{\textbf{z}}.\ \textbf{M}^{-1}.\tilde{\textbf{z}})}{\exp(\mathcal{L}(\boldsymbol{\theta}^{m-1})-\dfrac{1}{2}\textbf{z}^{0}.\ \textbf{M}^{-1}.\textbf{z}^{0})}$
       $\bigg\}$\\
       \hspace{1.5cm} $\boldsymbol{\theta}^{m}$ $\gets$ $\tilde{\boldsymbol{\theta}}$, $\textbf{z}^{m}$ $\gets$ -$\tilde{\textbf{z}}$
\EndIf  
\EndFor
\EndProcedure
\newline
\newline
\newline
\Function {Leapfrog-BurnIn}{$\tilde{\boldsymbol{\theta}}, \tilde{\textbf{z}}, \varepsilon, \textbf{B}$}
\State $\tilde{\textbf{z}} \gets \textbf{z}+(\varepsilon/2)\textbf{B}\nabla_{\boldsymbol{\theta}}\mathcal{L}(\boldsymbol{\theta})$
\State $\tilde{\boldsymbol{\theta}} \gets \boldsymbol{\theta}+\varepsilon\textbf{B}\tilde{\textbf{z}}$
\State $\tilde{\textbf{z}} \gets \textbf{z}+(\varepsilon/2)\textbf{B}\nabla_{\boldsymbol{\theta}}\mathcal{L}(\tilde{\boldsymbol{\theta}})$\\
\Return $\tilde{\boldsymbol{\theta}}$, $\tilde{\textbf{z}}$. 
\EndFunction
\newline
\Function {Leapfrog}{$\tilde{\boldsymbol{\theta}}, \tilde{\textbf{z}}, \varepsilon, \textbf{M}$}
\State $\tilde{\textbf{z}} \gets \textbf{z}+(\varepsilon/2)\nabla_{\boldsymbol{\theta}}\mathcal{L}(\boldsymbol{\theta})$
\State $\tilde{\boldsymbol{\theta}} \gets \boldsymbol{\theta}+\varepsilon\textbf{M}^{-1}\tilde{\textbf{z}}$
\State $\tilde{\textbf{z}} \gets \textbf{z}+(\varepsilon/2)\nabla_{\boldsymbol{\theta}}\mathcal{L}(\tilde{\boldsymbol{\theta}})$\\
\Return $\tilde{\boldsymbol{\theta}}$, $\tilde{\textbf{z}}$. 
\EndFunction
\end{algorithmic}
\end{algorithm}        
\section{NUMERICAL RESULTS}\label{section5}
In this section, four numerical examples are implemented to illustrate the efficiency of the proposed methods. In all examples, the tuning parameters ($\varepsilon$,$\tau$,$\sigma$) are systematically used as mentioned in \cref{section2,section4}. In the context of reliability problems, we use the default value $\tau=0.7$ as a starting point and then employ the ESJD metric \citep{wang2013adaptive} as described in \cref{section2}. The burn-in period is chosen to be on average 15\% of the total number of model calls, while the upper bound of the burn-in size is limited to 20\%. The described methods are compared to the Component-wise Metropolis-Hastings based Subset Simulation (CWMH-SuS). For the sake of comparison, we use two proposal distributions in CWMH-SuS, a uniform distribution of width 2 and a standard normal one. The parameters of Subset Simulation are chosen as $n_{s} =\,$1,000 and 2,000 for low- and high-dimensional simulations respectively, where $n_{s}$ is the number of samples for each subset level, and $p_{0} = 0.1$, where $p_{0}$ is the percentile of the samples that determines the intermediate subsets \citep{au2001estimation}. Comparisons are illustrated in terms of accuracy and computational cost. In particular, the tables show the \textit{P\textsubscript{F}} estimation, including the mean number of limit-state function calls in order to calculate the value and gradient of the target distribution, the analytical gradients are provided in all examples, in the HMCMC-based algorithms, and the value of the limit-state function in SuS. In all examples, the number of limit-state function evaluations for all methods has been set to be roughly the same to each other for comparison purposes. Results are based for all examples on 500 independently performed simulations, so that the sample mean and C.O.V of the results can be acquired. It should be noted that the ASTPA parameters are carefully chosen for all examples but are not optimized for any one. Hence, comparative and perhaps improved alternate performance might be achieved with a different set of parameters. 
\subsection{Example 1: parabolic/concave limit-state function}
The first example is expressed by the following limit state function for two standard
normal random variables \citep{der1998multiple}:
\begin{align}
g(\theta) = r - \theta_{2} - \kappa\ (\theta_{1}-e)^{2}
\end{align}   
where $r$, $\kappa$ and $e$ are deterministic parameters chosen as $r = 6$, $\kappa=0.3$ and $e = 0.1$. The probability of failure is 3.95E-5 and the limit-state function consists of two design points (failure modes), as seen in \cref{fig2}. For the HMCMC-based algorithms, the likelihood dispersion factor, $\sigma$, is 0.7 and the burn-in sample size is taken as 200. Consistent to the discussion in \cref{section4}, the trajectory length is set to $\tau=1$. \cref{tabel2} compares the number of model calls, the coefficient of variation and the $\mathop{\mathbb{E}}[\hat{P}_{F}]$ obtained by all tested methods. The Subset Simulation results are based on $n_{s} =\,$1,000. It is shown that the HMCMC approach gives significantly smaller C.O.V. than SuS and also outperforms it in terms of the $\mathop{\mathbb{E}}[\hat{P}_{F}]$. \cref{fig2} also demonstrates that the QNp-HMCMC samples accurately describe the two important failure regions.
\newcommand{\head}[1]{\textnormal{\textbf{#1}}}
\begin{table}[t!]
\caption{Performance of various methods for the parabolic/concave limit-state function}
\centering
\footnotesize
\setlength\tabcolsep{4pt}
\begin{tabular}{p{1cm}p{5cm}ccccccc}
  \toprule[1.5pt]
  \multirow{8}{*}{\shortstack[l]{$\sigma=0.7$ \\$\tau = 1$}} & 
  \multirow{2}{4cm}{\textbf{500 Independent Simulations}} & \multicolumn{2}{c}{\textbf{CWMH-SuS}} & \multicolumn{1}{c}{\head{HMCMC}} & \multicolumn{1}{c}{\head{QNp-HMCMC}}\\ 
  \cline{3-4}
  & & $U(-1,1)$ & $N(0,1)$\\ 
 \cmidrule(lr){2-6}
 &Number of model calls & 4,559 & 4,565 & 4,391 & 4,926 \\
 &C.O.V &0.62 & 0.65& 0.35 &0.39\\
 &$\mathop{\mathbb{E}}[\hat{P}_{F}]$ \ \ \ \ (Exact $P_{F}$ $\sim$ 3.95E-5) & 4.19E-5 & 4.14E-5 & 3.86E-5 &3.47E-5\\
 \bottomrule[1.5pt]
\end{tabular}\label{tabel2}
\end{table}

\subsection{Example 2: four-branch series system}
\begin{figure}[t!]
\centerline{\subfigure[]{\includegraphics[width=0.39\textwidth]{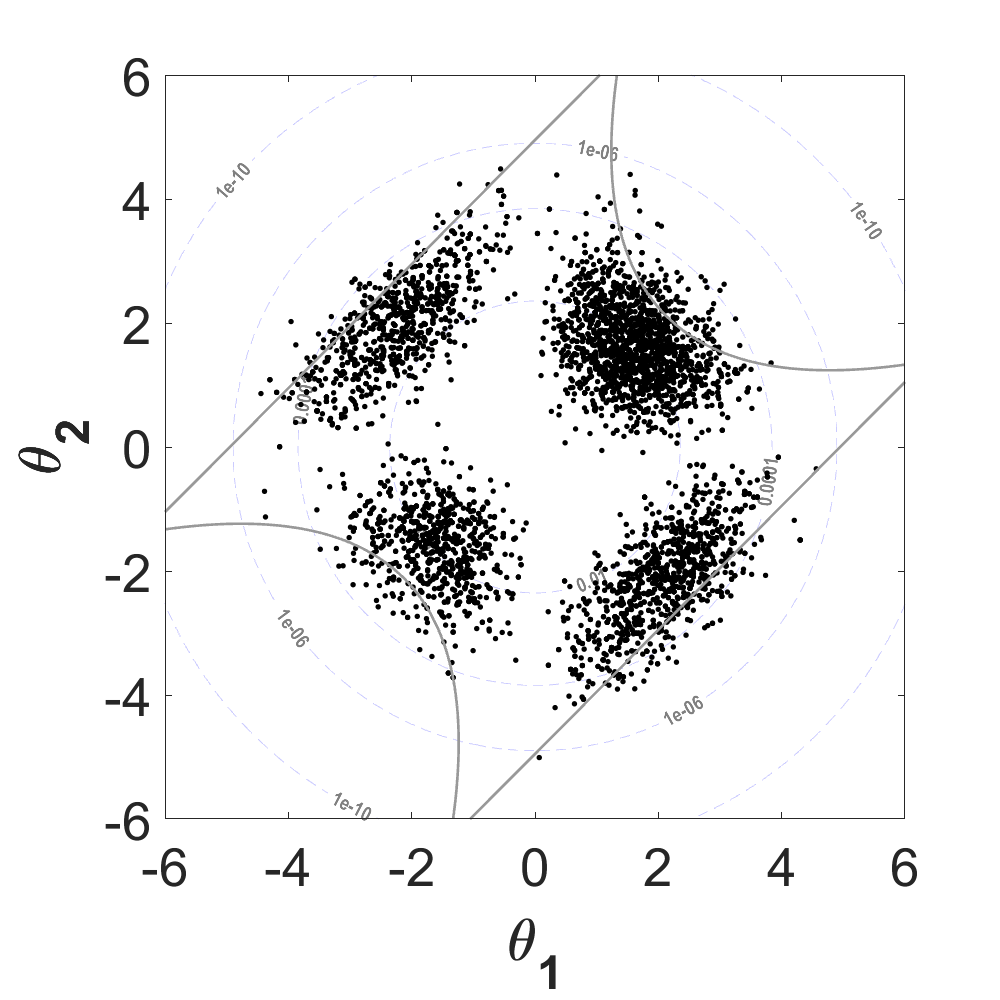}}
\quad
\subfigure[]{\includegraphics[width=0.39\textwidth]{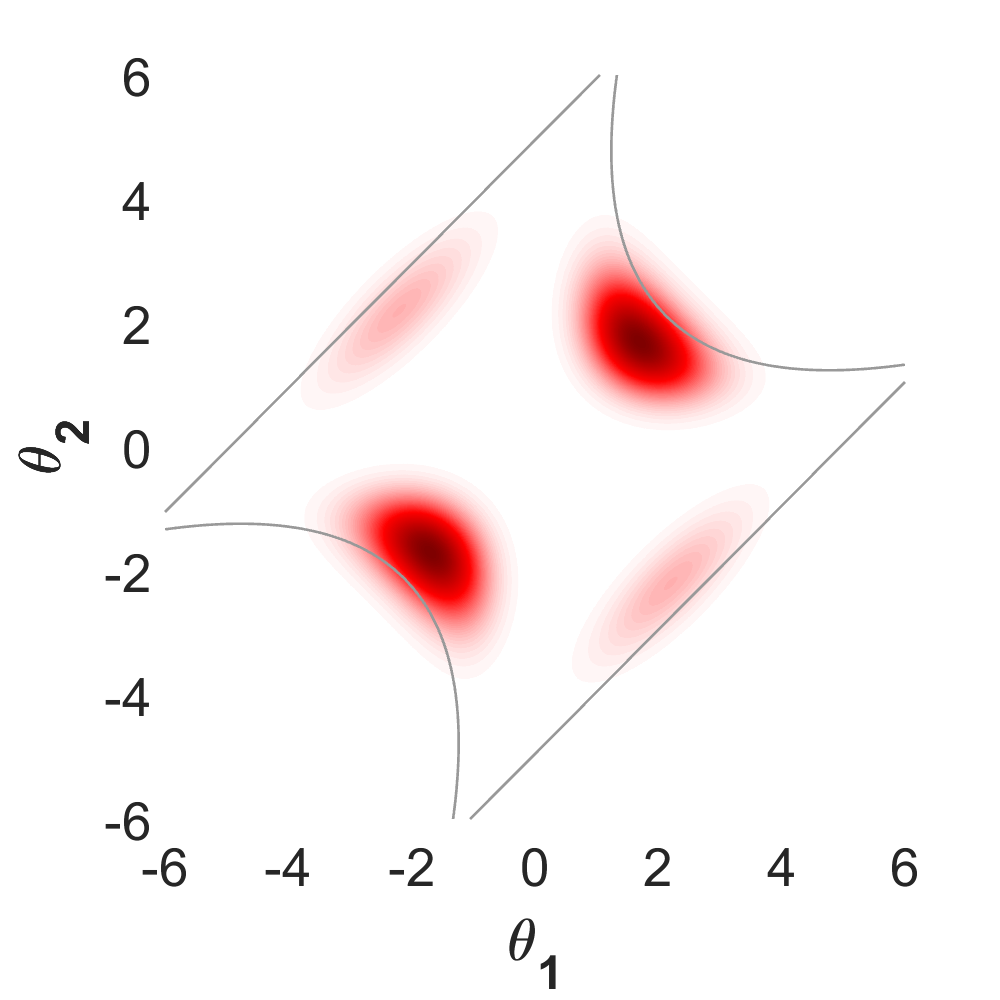}}}
\caption{(a) Simulated samples from the target distribution, (b) Analytical target distribution.}\label{fig8}
\end{figure}

This example is a well-known benchmark system reliability problem, defined by the following limit-state function in the standard normal space:
\begin{equation}
g(\boldsymbol{\theta})=\text{min}
\begin{cases}
3+ 0.1(\theta_{1} - \theta_{2})^{2} - (\theta_{1} - \theta_{2})/\sqrt{2}  \\
3+ 0.1(\theta_{1} - \theta_{2})^{2} + (\theta_{1} - \theta_{2})/\sqrt{2}  \\
(7/\sqrt{2})+ (\theta_{1} - \theta_{2}) \\
(7/\sqrt{2})+ (\theta_{2} - \theta_{1})
\end{cases}
\end{equation}
 The trajectory length is chosen as $\tau=1$ and the likelihood dispersion factor, $\sigma$, is fixed to 0.7. The burn-in is set to 200 samples. \cref{tabel4} shows that the SuS with uniform proposal gives more accurate $P_{F}$ estimation with smaller C.O.V than the HMCMC-based methods for the case of $n_{s}=\,$1,000. However, by increasing the sample size to $n_{s}=\,$2,000, it is seen that the HMCMC algorithm exhibits lower C.O.V compared to both SuS implementations. For the case of QNp-HMCMC, both the bias and C.O.V of the probability estimate considerably decrease with the sample size increase. \cref{fig8} shows the analytical target density of the four-branch limit-state function problem and samples from the target distribution using the HMCMC approach. As seen, the method achieves to efficiently sample all four important failure regions. 
\begin{table}[t!]
\caption{Performance of various methods for the four-branch series system}
\centering
\footnotesize
\setlength\tabcolsep{4pt}
\begin{tabular}{p{1.6cm}p{5cm}ccccccc}
  \toprule[1.5pt]
  \multirow{7}{*}{\shortstack[l]{$\sigma=0.7$ \\$\tau = 1$ \\ $n_{s}=\,$1,000}} & 
  \multirow{2}{5cm}{\textbf{500 Independent Simulations}} & \multicolumn{2}{c}{\textbf{CWMH-SuS}} & \multicolumn{1}{c}{\head{HMCMC}} & \multicolumn{1}{c}{\head{QNp-HMCMC}}\\ 
  \cline{3-4}
  & & $U(-1,1)$ & $N(0,1)$\\ 
 \cmidrule(lr){2-6}
 &Number of model calls & 2,841 & 2,852& 2,867&2,887\\
 &C.O.V &0.26 &0.30 &0.29&0.26 \\
 &$\mathop{\mathbb{E}}[\hat{P}_{F}]$ \ \ \  (Exact $P_{F}$ $\sim$ 2.20E-3) &2.23E-3  &2.26E-3&1.98E-3&1.91E-3 \\
 \bottomrule[1.5pt]
   \multirow{3}{*}{\shortstack[l]{$\sigma=0.7$ \\$\tau = 1$\\ $n_{s}=\,$2,000}}\rule{0pt}{2.5ex}
     &Number of model calls & 5,634 & 5,657 & 5,688&5,740\\
     &C.O.V &0.19 & 0.22& 0.13&0.17\\
     &$\mathop{\mathbb{E}}[\hat{P}_{F}]$ \ \ \  (Exact $P_{F}$ $\sim$ 2.20E-3) & 2.24E-3&2.23E-3  & 2.16E-3&2.11E-3 \\
    \bottomrule[1.5pt]
\end{tabular}\label{tabel4}
\end{table}

\subsection{Example 3: SDOF oscillator under impulse load}
In this example, a nonlinear undamped single-degree-of-freedom (SDOF) oscillator subjected to a rectangular impulse load is analysed, as described in \citep{bucher1990fast,schobi2017structural}. The limit-state function is given as:  
\begin{align}
g(k_{1},k_{2},M,r,T_{1},F_{1}) = 3r - \bigg\lvert \dfrac{2F_{1}}{M\omega_0^{2}} \sin(\frac{\omega_0 T_{1}}{2}) \bigg\rvert \label{eq26}
\end{align} 
where $\omega_0 = \sqrt{(k_{1}+k_{2})/M}$ is the natural frequency of the oscillator, $T_{1}$ is the duration of the impulse load, $M$ is the mass, $k_{1}$ and $k_{2}$ are the stiffnesses of the primary and secondary springs, $r$ is the displacement at which one of the springs yields, and $F_{1}$ is the amplitude of the force. The description of all random variables is listed in \cref{tabel15}. SuS results for both proposals are based on $n_{s} =\,$1,000. 
\begin{table}[t!]
\caption{Random variables of the undamped oscillator}
\centering
\small
\begin{tabular}{cccccccc}
  \toprule[1.5pt]
  Variable &Distribution  & Mean &C.O.V\\
  \hline
  $M$ & Gaussian  &1 & 0.05\\
  $k_{1}$ &Gaussian &1&0.1\\
  $k_{2}$ &Gaussian &0.1&0.1\\
  $r$ &Gaussian &0.5&0.1\\
  $T_{1}$ &Gaussian &1&0.2\\
  $F_{1}$ &Gaussian &0.6-0.45&$\frac{1}{6}$\\
 \bottomrule[1.5pt]
\end{tabular}\label{tabel15}
\end{table}
\begin{table}[t!]
\caption{Performance of various methods for the undamped oscillator example}
\centering
\footnotesize
\setlength\tabcolsep{4pt}
\begin{tabular}{p{1.5cm}p{5cm}ccccccc}
  \toprule[1.5pt]
  \multirow{8}{*}{\shortstack[l]{$\sigma=0.1$ \\$\tau = 0.7$\\$\mu_{F_{1}}=0.6$}} & 
  \multirow{2}{5cm}{\textbf{500 Independent Simulations}} & \multicolumn{2}{c}{\textbf{CWMH-SuS}}& \multicolumn{1}{c}{\head{HMCMC}} & \multicolumn{1}{c}{\head{QNp-HMCMC}}\\ 
  \cline{3-4}
  & & $U(-1,1)$ & $N(0,1)$\\ 
 \cmidrule(lr){2-6}
 &Number of model calls & 5,170 & 5,160 & 5,132 &5,119\\
 &C.O.V &0.67 & 0.51 &0.14&0.11\\
 &$\mathop{\mathbb{E}}[\hat{P}_{F}]$ \ \ \ \ (Exact $P_{F}$ $\sim$ 9.09E-6) & 9.68E-6 & 9.55E-6& 9.10E-6&9.08E-6\\
 \bottomrule[1.5pt]
    \multirow{3}{*}{\shortstack[l]{$\sigma=0.1$ \\$\tau = 0.7$\\$\mu_{F_{1}}=0.45$}}\rule{0pt}{2.5ex}
      &Number of model calls & 7,583 & 7,617& 7,523&7,515\\
      &C.O.V &0.77 & 0.70& 0.21&0.15\\
      &$\mathop{\mathbb{E}}[\hat{P}_{F}]$ \ \ \ \ (Exact $P_{F}$ $\sim$ 1.55E-8) & 1.67E-8&1.50E-8  & 1.52E-8&1.51E-8 \\
     \bottomrule[1.5pt]
\end{tabular}\label{tabel16}
\end{table}
All variables are first transformed to the standard normal space. Results are shown in \cref{tabel16} for two cases, by changing the mean value, $\mu_{F_{1}}$, of $F_{1}$. For the HMCMC-based methods, the trajectory length and the likelihood dispersion factor are chosen as $\tau = 0.7$ and $\sigma=0.1$ respectively. The burn-in sample size is set to 500. It is shown in this example that the QNp-HMCMC approach provides significantly more accurate and stable results in terms of the C.O.V. and $\mathop{\mathbb{E}}[\hat{P}_{F}]$. Particularly for the lowest failure probability level, QNp-HMCMC approach noticeably outperforms all other methods. As results indicate, the QNp-HMCMC method is roughly insensitive to the failure probability level and there is no negative influence on the method when changing $\mu_{F_{1}}$. For the two SuS variants, it is noteworthy to say here that the SuS with the standard normal proposal distribution indicates reasonably better performance in this example than the one with the uniform proposal.   
\subsection{Example 4: SDOF oscillator under white noise excitation}
In this last example, we consider a SDOF oscillator, initially at rest, with natural frequency $\omega = 7.85\, rad/s$ and damping ratio $\xi = 0.02$, subjected to a Gaussian white noise ($W(t)$) excitation with spectral density of magnitude $S_{0}=1$.
The response of the system is computed at discrete time instants $\{t_{j} = (j-1)\Delta t:j = 1,...n\}$ with $\Delta t=0.05$, and the duration of study is $T = 5\, sec$. Thus, the number of time instants is equal to $n=T/\Delta t + 1 = 101$. The state vector $\boldsymbol\theta$ is the sequence of i.i.d. standard normal random variables that generate the $W(t_{j}) = \sqrt{\frac{2\pi S_{0}}{\Delta t}}\theta_{j}$ at the discrete time instants, resulting in $101$ involved random variables in this example. Failure is characterized by the positive displacement response exceeding a threshold level $R$: $g(\boldsymbol\theta)= R - max\{Y(t)\}$.\par  
\begin{table}[t!]
\caption{Performance of various methods for SDOF oscillator under white noise}
\centering
\footnotesize
\setlength\tabcolsep{4pt}
\begin{tabular}{p{1.5cm}p{5cm}ccccccc}
  \toprule[1.5pt]
  \multirow{8}{*}{\shortstack[l]{$\sigma=0.2$ \\$\tau = 0.9$\\$R=1.8$}} & 
  \multirow{2}{5cm}{\textbf{500 Independent Simulations}} & \multicolumn{2}{c}{\textbf{CWMH-SuS}}& \multicolumn{1}{c}{\head{HMCMC}} & \multicolumn{1}{c}{\head{QNp-HMCMC}}\\ 
  \cline{3-4}
  & & $U(-1,1)$ & $N(0,1)$\\ 
 \cmidrule(lr){2-6}
 &Number of model calls & 11,000 & 11,011 & 11,063 &11,059\\
 &C.O.V &0.32 & 0.35 &0.30&0.24\\
 &$\mathop{\mathbb{E}}[\hat{P}_{F}]$ \ \ \ \ (Exact $P_{F}$ $\sim$ 2.53E-6) & 2.58E-6 & 2.63E-6& 2.57E-6&2.55E-6\\
 \bottomrule[1.5pt]
    \multirow{3}{*}{\shortstack[l]{$\sigma=0.2$ \\$\tau = 0.9$\\$R=2$}}\rule{0pt}{2.5ex}
      &Number of model calls & 13,578 & 13,646& 13,644&13,618\\
      &C.O.V &0.41 & 0.48& 0.34&0.29\\
      &$\mathop{\mathbb{E}}[\hat{P}_{F}]$ \ \ \ \ (Exact $P_{F}$ $\sim$ 1.11E-7) & 1.16E-7&1.14E-7  & 1.13E-7&1.12E-7 \\
     \bottomrule[1.5pt]
\end{tabular}\label{tabel17}
\end{table}
The burn-in sample size is taken as 1,000 for the HMCMC-based methods. SuS results are based on $n_{s} =\,$2,000. It is seen in \cref{tabel17} that the QNp-HMCMC approach shows more accurate and efficient results in terms of C.O.V. and $\mathop{\mathbb{E}}[\hat{P}_{F}]$. Compared to the HMCMC approach, this example agrees with the additional results in \cite{Nikbakht2019HMCMC} and confirms that the application of QNp-HMCMC in high-dimensional reliability problems is in general more attractive. By decreasing the target failure probability, results also reveal that QNp-HMCMC gives us a substantially improved estimation in comparison to all other methods.            
\section{CONCLUSIONS}\label{section6}
A novel approach for estimation of rare event probabilities termed Approximate Sampling Target with Post-processing Adjustment (ASTPA), is presented in this paper,  suitable for low- and high-dimensional problems, very small probabilities and multiple failure modes. ASTPA can provide an accurate unbiased estimation of the failure probabilities with an efficient number of limit-state function evaluations. The basic idea of ASTPA is to construct a relevant target distribution by weighting the high-dimensional random variable space through a one-dimensional likelihood model, using the limit-state function. To sample from this target distribution we utilize gradient-based HMCMC schemes, including our newly developed Quasi-Newton based mass preconditioned HMCMC algorithm (QNp-HMCMC) that can sample very adeptly, particularly in difficult cases with high-dimensionality and very small failure probabilities. Finally, an original post-sampling step is also devised, using an inverse importance sampling procedure based on the samples. The performance of the proposed methodology is examined and compared very successfully herein against Subset Simulation in a series of static and dynamic low- and high-dimensional benchmark problems. As a general guideline, QNp-HMCMC is recommended to be used for problems with more than 20 dimensions, where traditional HMCMC schemes may not perform that well. However, even in lower dimensions QNp-HMCMC performs reasonably well and is still a competitive algorithm. Since we are utilizing gradient-based sampling methods, all of our analyses and results are based on the fact that analytical gradients can be computed. In cases where numerical schemes are needed for the gradient evaluations, then HMCMC methods will not be competitive in relation to SuS. It should also be pointed out that different combinations of the HMCMC and QNp-HMCMC algorithms can be possible, based on problem-specific characteristics. Some of the ongoing and future work is directed towards exploring various ASTPA variants, and on estimating first-passage problems under numerous settings and high-dimensional parameter spaces.
\section*{REFERENCES}
\renewcommand\refname{}
\vspace*{-0.5cm}
\bibliographystyle{unsrt}
\bibliography{references1}  
\end{document}